\documentclass[preprint,12pt]{elsarticle}

\usepackage{amssymb}
\usepackage{amsmath}

\journal{Acta Astronautica}

\begin{document}

\begin{frontmatter}



\title{Forecasting Thermoacoustic Instabilities in Liquid Propellant Rocket Engines Using Multimodal Bayesian Deep Learning}


\author[inst1]{Ushnish~Sengupta}

\affiliation[inst1]{organization={Department of Engineering},
            addressline={University of Cambridge}, 
            city={Cambridge},
            state = {Cambridgeshire},
            postcode={CB2~1PZ},
            country={United Kingdom}}

\author[inst2]{Günther~Waxenegger-Wilfing}

\affiliation[inst2]{organization={Institute of Space Propulsion},
            addressline={German Aerospace Center (DLR)}, 
            city={Hardthausen},
            state={Baden-Württemberg},
            postcode={74239},
            country={Germany}}

\author[inst2]{Jan Martin}
\author[inst2]{Justin~Hardi}
\author[inst1]{Matthew~P.~Juniper \corref{cor1}}
\cortext[cor1]{Corresponding author. Email address: mpj1001@cam.ac.uk.}
\begin{abstract}
The 100 MW cryogenic liquid oxygen/hydrogen multi-injector combustor BKD operated by the DLR Institute of Space Propulsion is a research platform that allows the study of thermoacoustic instabilities under realistic conditions, representative of small upper stage rocket engines. We use data from BKD experimental campaigns in which the static chamber pressure and fuel-oxidizer ratio are varied such that the first tangential mode of the combustor is excited under some conditions. We train an autoregressive Bayesian neural network model to forecast the amplitude of the dynamic pressure time series, inputting multiple sensor measurements (injector pressure/ temperature measurements, static chamber pressure, high-frequency dynamic pressure measurements, high-frequency OH* chemiluminescence measurements) and future flow rate control signals. The Bayesian nature of our algorithms allows us to work with a dataset whose size is restricted by the expense of each experimental run, without making overconfident extrapolations. We find that the networks are able to accurately forecast the evolution of the pressure amplitude and anticipate instability events on unseen experimental runs 500 milliseconds in advance. We compare the predictive accuracy of multiple models using different combinations of sensor inputs. We find that the high-frequency dynamic pressure signal is particularly informative. We also use the technique of integrated gradients to interpret the influence of different sensor inputs on the model prediction. The negative log-likelihood of data points in the test dataset indicates that predictive uncertainties are well-characterized by our Bayesian model and simulating a sensor failure event results as expected in a dramatic increase in the epistemic component of the uncertainty.
\end{abstract}




\end{frontmatter}


\section{Introduction}
\label{sec:sample1}
High-frequency thermoacoustic instabilities are a potentially catastrophic phenomenon in rocket and aircraft engines. They usually arise from the coupling between unsteady heat release rate and acoustic eigenmodes of the combustion chamber. Acoustic waves incident on a flame cause fluctuations in its heat release rate. If a flame releases more (less) heat than average during instants of higher (lower) local pressure, then more work is done by the gas during the acoustic expansion phase than is done on it during the acoustic compression phase. If this work is not dissipated then the oscillation amplitude grows and the system becomes thermoacoustically unstable. This was formalized by Chu (Equation 26 in \cite{chu1965energy}) who formulated the following criterion for instability:
\begin{equation}
    \int_{0}^{\tau}\int_{0}^{V}p'(x,t)q'(x,t)dxdt > \frac{p_0 \gamma}{\gamma - 1} \int_{0}^{\tau}\int_{0}^{V} \phi(x,t) dxdt
\end{equation}
where $p'$ and $q'$, are the perturbations in pressure and the heat release, respectively, $p_0$ is the pressure of the unperturbed system, $\tau$, $V$ and $\phi$ are the period of oscillation, the combustor volume and the viscous dissipation function, respectively.

Thermoacoustic instabilities can be challenging to model and predict because they are usually sensitive to small changes in a combustion system \cite{juniper2018sensitivity}. The large pressure fluctuations and accelerated heat transfer stemming from these oscillations can lead to structural failure, because rocket engines have high energy densities and low factors of safety in their structural design. From the F1 rocket engine, which underwent costly redesigns in the 1960s \cite{oefelein1993comprehensive} to the modern Japanese LE-9 engine \cite{watanabe2016combustion}, they continue to plague the development of liquid propellant rocket engines.

Designers can suppress thermoacoustic oscillations in three major ways. The first approach is to fit passive devices such as acoustic liners \cite{oefelein1993comprehensive}, baffles \cite{eldredge2003absorption} or resonators \cite{zinn1970theoretical}. The second is tuned passive control, where sensor measurements are used to diagnose the onset of thermoacoustic instabilities which can then be avoided by adjusting a passive device or the operating point \cite{kobayashi2019early}. The third is active feedback control \cite{dowling2005feedback}. The challenging demands of high power output, reliability and robustness placed on an active feedback control system means that feedback control is not used in rocket engines.

The construction of instability precursors from pressure measurements and optical measurements has attracted considerable attention from the combustion research community \cite{Juniper2018}. Lieuwen \cite{Lieuwen2005} used the autocorrelation decay of combustion noise, filtered around an acoustic eigenfrequency, to obtain an effective damping coefficient for the corresponding instability mode. Other researchers have used tools from nonlinear time series analysis, which capture the transition from the chaotic behaviour displayed by stable turbulent combustors to the deterministic acoustics during instability, such as Gottwald's 0-1 test \cite{Nair2013} and the Wayland test for nonlinear determinism \cite{Gotoda2011} to assess the stability margin of a combustor. Nair \textit{et al} \cite{Nair2014intermittency} reported that instability is often presaged by intermittent bursts of high-amplitude oscillations and used recurrence quantification analysis (RQA) to detect these. In a later paper \cite{Nair2014}, Nair and coworkers noted that combustion noise tends to lose its multifractality as the system transitions to instability. They proposed the Hurst exponent as an indicator of impending instability. Measures derived from symbolic time series analysis (STSA) \cite{sarkar2016dynamic} and complex networks \cite{murugesan2016detecting} are also able to capture the onset of instability.

Recent studies have explored machine learning techniques to learn precursors of instability from data. These promise greater accuracy than physics-based precursors of instability, though at the cost of robustness and generalizability. Hidden Markov models constructed from the output of STSA \cite{jha2018markov} or directly from pressure measurements \cite{mondal2018early} have been used to classify the state of combustors. Hachijo \textit{et al} \cite{Hachijo2019} have projected pressure time series onto the entropy-complexity plane and used support vector machines (SVMs) to predict thermoacoustic instability. SVMs were also employed by Kobayashi \textit{et al} \cite{Kobayashi2019}, who used them in combination with principal component analysis and ordinal pattern transition networks to build precursors from simultaneous pressure and chemiluminiscence measurements. Sengupta \textit{et al} \cite{sengupta2020bayesian} showed that the power spectrum of the noise can be used to predict the linear stability of a thermoacoustic eigenmode using Bayesian neural networks. Related work by McCartney \textit{et al} \cite{mccartney2020} used the detrended fluctuation analysis (DFA) spectrum of the pressure signal as input to a random forest and finds that this approach compares favorably to precursors from the literature. A recent study by Gangopadhyay and co-workers \cite{GANGOPADHYAY2021100067} trained a 3D Convolutional Selective Autoencoder on flame videos to detect instabilities. Transfer learning is also being explored \cite{MONDAL2021100085} as a means of efficiently transferring knowledge of precursors across machines.

Most papers in the literature train indicators of approaching instability. In this study, on the other hand, we use nonlinear autoregressive time series modeling with a Bayesian Neural Network to forecast, with uncertainties, the future amplitude of pressure fluctuations given the history of sensor signals and future flow rate control signals. This informs the engine controller of the expected timing and potential severity of an instability event. This approach also lets us take advantage of the rich instrumentation of our rig by integrating multimodal sensor data into our model: high frequency dynamic pressure ($p'(t)$) and OH* chemiluminescence ($I'(t)$) measurements, injector pressures ($p_{H2}$, $p_{O2}$) and temperatures ($T_{H2}$, $T_{O2}$), chamber static pressure ($p_{\text{cc}}$) and fuel flow rates ($\dot m_{O2}, \dot m_{H2}$). We compare models with different combinations of sensors, different sensor-derived features and different lengths of signal histories in the inputs. We find that, although operating parameters and mass flow rate control signals can anticipate instabilities on their own, inclusion of dynamic pressure signal history, in particular, results in a measurable increase in the forecast accuracy. It is not possible to predict triggered thermoacoustic instabilities from operating parameters and control signals alone, so the need to integrate multimodal sensor data with operating parameters and control signals in an instability forecasting framework is exigent.

The size of our dataset is limited by the high cost and effort required for each additional test run on the rocket thrust chamber. Fortunately, the uncertainty-aware nature of the Bayesian Neural Network allows us to work in this medium-data regime without overconfident extrapolation. We compute log-likelihoods on the test dataset and find that the uncertainty is well-quantified. We also simulate a sensor failure event to assess the robustness of our Bayesian forecasting tool. As expected, the network exhibits large epistemic uncertainties during the failure event, signalling the unfamiliar out-of-distribution nature of the inputs. Finally we use the interpretation technique of integrated gradients (IG) to understand how the different input features in our model influence the predictions. 

\section{Experimental Setup}

The dataset used in this study is derived from experiments performed on the research combustor BKD \cite{suslov2003test} \cite{sender2016l42} operated at the P8 test facility \cite{frohlke1997first} \cite{koschel1996p8} of the DLR Institute of Space Propulsion in Lampoldshausen. It has three main components: an injector head, a cylindrical combustion chamber and a convergent-divergent nozzle, as shown in Figure 1. The combustion chamber is water-cooled and is designed to deal with the high thermal loads that are expected during instability events. The L42 injector head has 42 shear coaxial injectors and is operated with a liquid oxygen (LOX)/ hydrogen (LH2) propellant combination. The cylindrical combustion chamber is 80 mm in diameter and the nozzle throat diameter is 50 mm, resulting in a contraction ratio of 2.56.  The chamber static pressure ($p_{\text{cc}}$) is varied between $50 - 80$~bar and mixture ratio of oxidizer to fuel (ROF = $\dot m_{O2}/\dot m_{H2}$) between $2 - 6$. The experiments considered here have a LOX injection temperature $T_{O2}$ around 110~K and a hydrogen injection temperature $T_{H2}$ around 100~K.

For the operating point with $p_{\text{cc}}=80$ bar, ROF $=6$, the total propellant mass flow rate is 6.7 kg/s, the theoretical thermal power is 100~MW and the thrust achieved is about 24~kN. These specifications place BKD at the lower end of small upper stage engines.
\begin{figure}
\centering
\includegraphics[scale = 0.375]{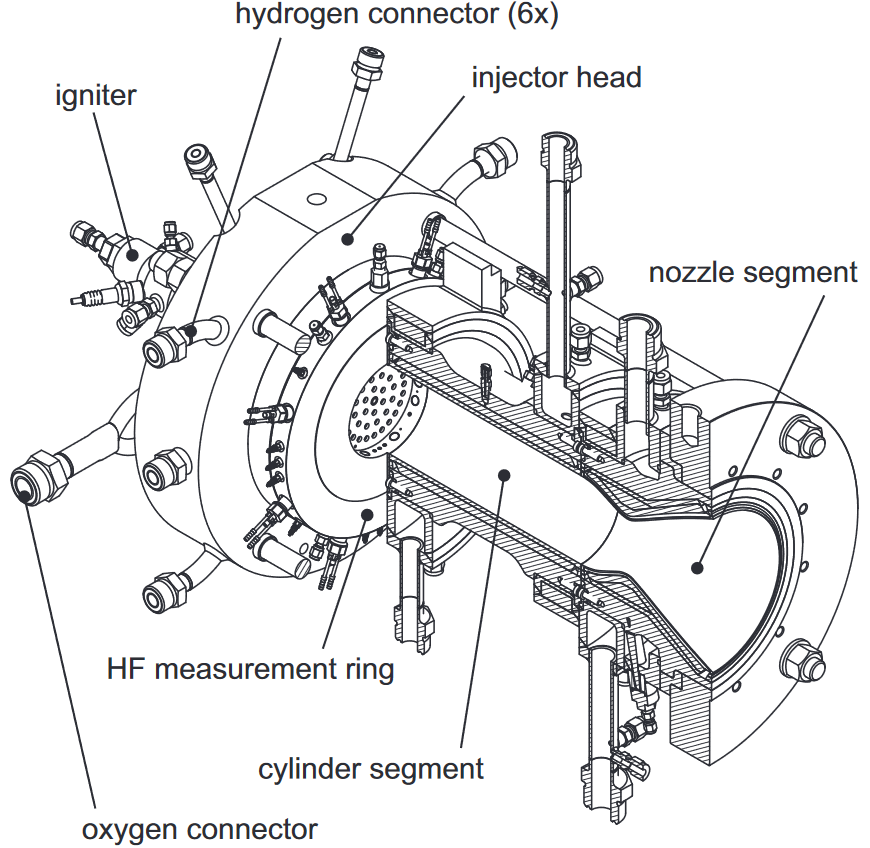}
\caption{\label{fig_bkd} Experimental thrust chamber BKD.}
\end{figure}

\begin{figure}
\centering
\includegraphics[scale = 0.5]{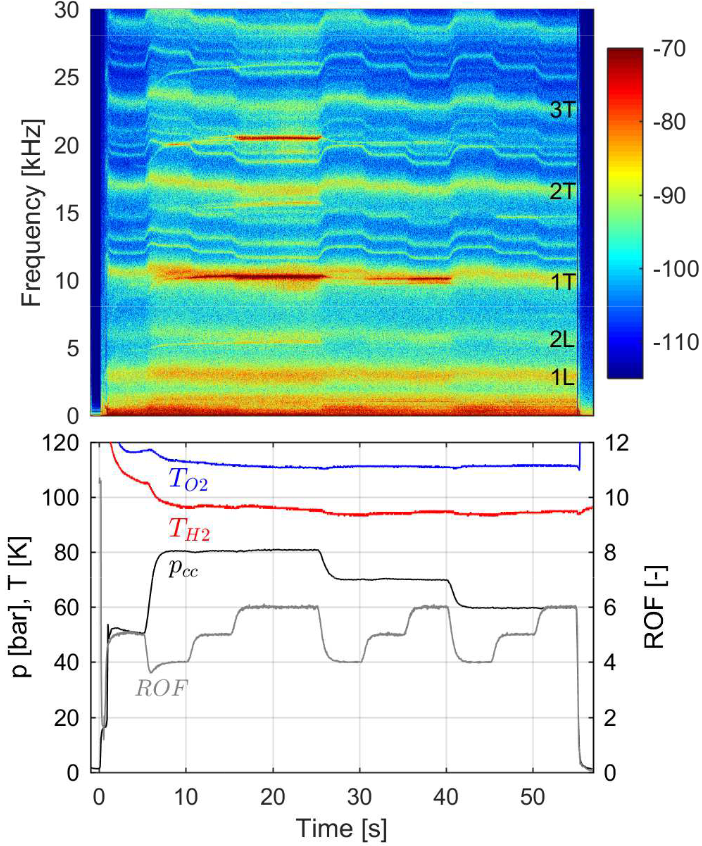}
\caption{\label{fig_Spectrogram_Type1} BKD test sequence and $p'$ spectrogram showing self-excited instability of the first tangential mode \cite{SGJPP}. $p_{\text{cc}}$ is the chamber static pressure, ROF is the mixture ratio of oxidizer to fuel, and $T_{H2}, T_{O2}$ are the LOX and hydrogen injection temperatures respectively.}
\end{figure}

A representative BKD test sequence, along with a spectrogram of dynamic pressure oscillations inside the combustion chamber, is shown in Figure \ref{fig_Spectrogram_Type1}. Stable and unstable operating conditions can be identified in the spectrogram. Strong high-frequency combustion instabilities of the first tangential (1T) mode at about 10~kHz were excited consistently when the operating point at $p_{\text{cc}} = $ 80 bar, ROF $=6$ was approached. The coupling mechanism for this instability is injection-driven \cite{SGJPP}. The flame dynamics are modulated by the LOX post acoustics and combustion instabilities emerge when the frequency of the 1T chamber mode matches the second longitudinal eigenmodes of the LOX posts. This mechanism has been confirmed using high-speed flame imaging \cite{Armbruster19JPP}.

The extreme conditions within the thrust chamber, where temperatures reach 3600~K and the pressure reaches 80 bar, limit the diagnostic instruments for our instability investigations. A specially designed measurement ring is placed between the injector head and the cylindrical combustion chamber segment, as shown in Figure \ref{fig_bkd}. At this location, the temperatures are moderated by the injection of cryogenic propellants, which allows the mounted sensors to survive several test runs. Eight Kistler type 6043A water-cooled high-frequency piezoelectric pressure sensors are flush-mounted in the ring with an even circumferential distribution, in order to measure the chamber pressure oscillations $p'(t)$. The high-frequency pressure sensors have a measurement range set to $\pm30$~bar and a sampling rate of 100~kHz. An anti-aliasing filter with a cutoff frequency of 30~kHz is applied. Three fibre-optical probes are used to record the OH* radiation intensity $I'(t)$ of the individual flames. A small sapphire rod in the probe creates optical access. The full acceptance angle of the optical probes is approximately $2 ^{\circ}$. In order to capture the OH* radiation signal, the probes are equipped with inteference filters with a center wavelength of 310 nm. The sampling frequency of the $I'(t)$ signals is also 100~kHz. There are also several low sampling frequency (100 Hz) sensors that measure the chamber static pressure $p_{\text{cc}}$, the mass flow rates of fuel $\dot m_{H2}$ and oxidizer $\dot m_{O2}$ and injector temperatures $T_{H2}, T_{O2}$.

\section{Methods}
\label{sec:methods}

\subsection{Nonlinear autoregressive models with exogenous variables (NARX) for timeseries modeling}
To forecast instabilities, we train NARX-style models \cite{548162} to capture the functional relationship between the history of the sensor data, control signals and future pressure fluctuations. These model the evolution of pressure fluctuations as follows:

\begin{equation}
    \text{log}(p'_{\text{rms}}(t_m+\Delta t)) = f(\mathbf{x}_{m}, \mathbf{x}_{m-1}, ... \mathbf{x}_{m-h}, \mathbf{u}_{m+f}) + \sigma_{\epsilon}
\end{equation}

Here, $p'_{\text{rms}}(t_m+\Delta t)$ is the root mean square amplitude of the dynamic pressure fluctuation signal in a 50 ms window centered around $t_m + \Delta t$. We choose $\Delta t = 500$ ms for our models, because we consider this to be sufficient warning for a rocket engine controller to react and take appropriate action. The state variables $\mathbf{x}_{m-i}$ consist of various sensor data (or transformations thereof) at times $t_m - 50\times i$ ms.  The order of dependence $h$ determines how far in the past we look. In this study, we explore models with $h = 0$ and $h = 2$. The elements of $\mathbf{u}_{m+f}$ are the two mass flux control signals $\dot m_{H2, set}$ and $\dot m_{O2, set}$ averaged over the window between $t_m$ and $t_m+\Delta t$. These are the exogenous variables whose future values are known and can therefore be included in the inputs. It is possible to train models that capture higher order dependence on instantaneous values of the control signals instead of reducing them to an average, but we find that this does not improve predictive accuracy in our problem. $\sigma_{\epsilon}$ is the homoskedastic (constant) observation noise or aleatoric uncertainty term, which is assumed to be Gaussian. We forecast the logarithm of $p'_{\text{rms}}(t_m+\Delta t)$ instead of $p'_{\text{rms}}(t_m+\Delta t)$ because the noise in the pressure fluctuations increases with the amplitude, and so the simplifying assumption of constant homoskedastic noise only holds approximately once this logarithmic transform has been applied. It also forces $p'_{\text{rms}}$ predictions to be positive.

We model the nonlinear function $f$ in Equation (2) as a feedforward neural network with a simple multi-layer perceptron (MLP) architecture. In a standard neural network, one obtains a point estimate of the network parameters. Here, however, we train a machine learning algorithm on a dataset whose size is limited by the cost of performing experiments on large rocket thrust chambers. With a standard neural network this size limit could lead to naive and inaccurate extrapolations when the network encounters out-of-distribution inputs that are different from those which were experienced during training. In safety-critical systems such as rockets, control decisions based on overconfident predictions could result in dire consequences. Bayesian Neural Networks (BNNs) are an elegant solution to this problem because they provide a confidence interval as well as a prediction.

\subsection{Bayesian Neural Network Ensembles}
The Bayesian framework for training neural networks \cite{Neal1996} involves placing a sensible prior probability distribution $p(\theta)$ over the parameters $\theta$ of the network. Observations $\mathcal{D}$ are then used to update our belief about the parameter distribution. The posterior distribution $p(\theta|\mathcal{D})$ is obtained using Bayes' rule.

\begin{equation}
p(\theta|\mathcal{D}) = \frac{p(\mathcal{D}|\theta)p(\theta)}{\int{p(\mathcal{D}|\theta)p(\theta)d\theta}}
\end{equation}

Bayesian models are resistant to overfitting on small datasets and uncertainty aware. They are also useful for continual learning throughout the lifetime of a device in operation \cite{Li2019}. Exact Bayesian inference, however, is not possible in neural network models because the integral $\int{p(\mathcal{D}|\theta)p(\theta)d\theta}$ is analytically intractable. The most widely used numerical method to integrate over the posterior, Markov Chain Monte Carlo, is too computationally expensive to be practical. Variational inference is often used \cite{Blundell2015}, in which the posterior is approximated using a convenient parametrization and the Kullback-Leibler divergence between this variational distribution and the true posterior is minimized using backpropagation. However, although computationally cheap, mean-field variational inference also has drawbacks such as not capturing correlations between parameters. Recently, a different approximate inference method, based on ensembling, has been proposed. This is called the anchored ensembling algorithm. It is cheap, simple and scalable but manages to outperform variational inference in several uncertainty quantification benchmarks \cite{Pearce2018}.

Consider a dataset of $N_{D}$ data points ${(\mathbf{x}_n, y_n)}$, where each data point consists of features $\mathbf{x}_n\in{\mathbb{R}}^D$ and output $y_n\in{\mathbb{R}}$. Define the likelihood for each data point as $ p(y_n \mid \mathbf{\theta}, \mathbf{x}_n, \sigma_{\epsilon}^2) = \mathcal{N}(y_n \mid \mathrm{NN}(\mathbf{x}_n;\mathbf{\theta}), \sigma_{\epsilon}^2)$, where $\mathrm{NN}$ is a neural network whose weights and biases form the latent variables $\mathbf{\theta}$ while $\sigma_\epsilon^2$ is the observation noise. Define the prior on the weights and biases $\mathbf{\theta}$ to be the standard normal $p(\mathbf{\theta}) = \mathcal{N}(\mathbf{\theta} \mid \mathbf{\mu}_{prior}, {\Sigma}_{prior}).$ The anchored ensembling algorithm then does the following:
\begin{enumerate}
    \item The parameters $\theta_{anc,j}$ of each $j$-th member of our neural network ensemble are initialized by drawing from the prior distribution $\mathcal{N}(\mathbf{\mu}_{prior}, {\Sigma}_{prior})$.
    \item Each ensemble member is trained using a modified loss function that anchors the parameters to their initial values. The loss function for the $j$-th ensemble member is given by $Loss_{j} = \sum_{i=1}^{N_{D}}(y_i-\hat{y}_j(\mathbf{x}_i,t_i))^2 +  \|  \boldsymbol{\Sigma}_{prior}^{-1/2}({\theta}_j - {\theta}_{anc, j})\|_2^{2}$, where $\hat{y}_j(\mathbf{x}_i,t_i)$ is the neural network output and the $i$-th diagonal element of ${\Sigma}$ is the ratio of data noise to the prior variance of the $i$-th parameter.
\end{enumerate}

Pearce \textit{et al} \cite{Pearce2018} prove that this procedure approximates the true posterior distribution for wide neural networks and that the trained neural networks in the ensemble may be treated as samples from an approximate posterior distribution. The resulting ensemble has predictions that converge when they are well-supported by the training data and diverge when they are not. The variance of the ensemble predictions is an estimate of the Bayesian model's epistemic uncertainty. Epistemic or knowledge uncertainty is the reducible portion of uncertainty which stems from uncertainty in model parameters and decreases as more data is added. The total uncertainty of each prediction is obtained by adding the epistemic uncertainty and the irreducible aleatoric uncertainty $\sigma_{\epsilon}$ which stems from observation noise:
\begin{equation}
    \sigma^2_{\text{total}} = \sigma^2_{\epsilon} + \frac{1}{N_e}\sum_j \hat{y}^2_j - (\frac{1}{N_e}\sum_j \hat{y}_j)^2
\end{equation}
\begin{figure}
\centering
\includegraphics[scale = 0.25]{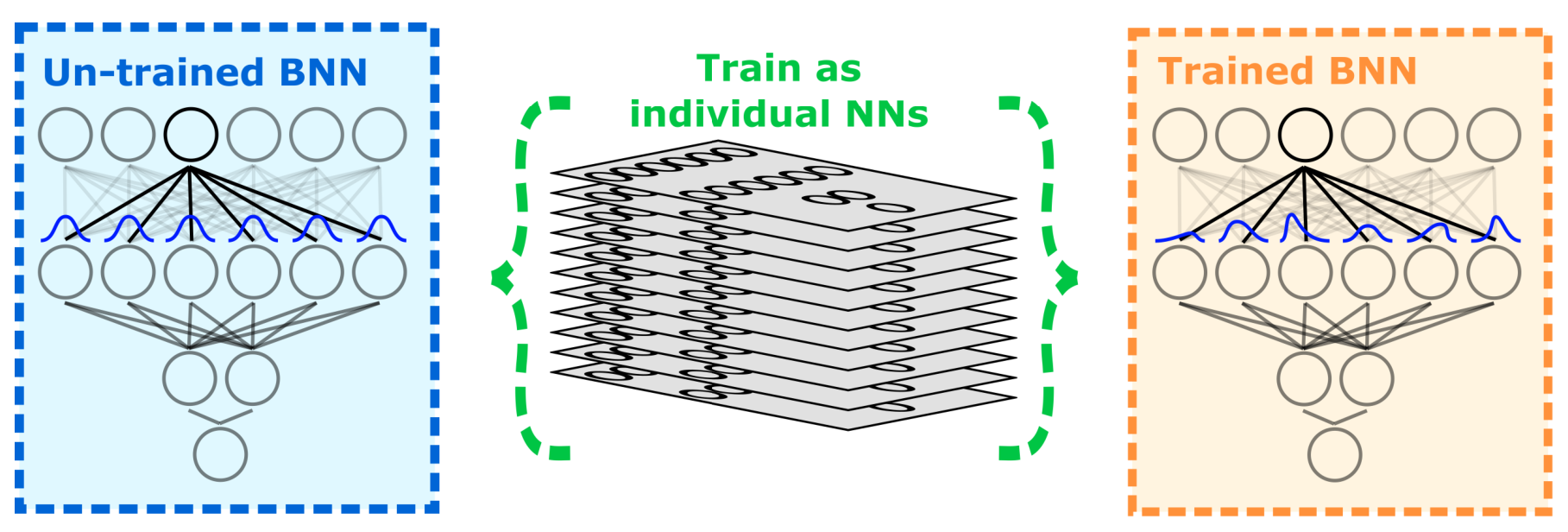}
\caption{\label{BNN} Approximate Bayesian inference using ensembles of neural networks.}
\end{figure}

\subsection{Interpretation using Integrated Gradients}
Neural networks have a reputation as black boxes, which disincentivizes their application to cases in which practitioners must understand why the algorithm is or is not working. We use the technique of integrated gradients (IG) \cite{sundararajan2017axiomatic} to attribute the predictions of our network ensemble to the input features. This is a simple scalable method that only requires access to the gradient operation. IG has several desirable properties that other attribution methods lack, such as implementation invariance, sensitivity, linearity, completeness and preservation of symmetry.

For deep neural networks, the gradients of the output with respect to the input are an analog of the linear regression coefficients.  In a linear model, the regression coefficients characterize the relationship between input and output variables globally.  In a nonlinear neural network, however, the gradient of the output at a point merely characterizes the local relationship between a predictor variable and the output.  IG computes the path integral of the gradient of the outputs with respect to the inputs around a baseline to the input under consideration.  For an image recognition algorithm, a completely black image could be a reasonable choice of baseline, while for a regression problem like ours, where the input variables were normalized to have zero mean and unit standard deviation, the average input of all zeros is a sensible baseline.  We consider the straight-line path (in the feature space $\mathbb{R}^p$) from the baseline $\boldsymbol{x}'$ to the input $\boldsymbol{x}$ being considered. The IG attribution $attr_{j}$ for the $j$-th feature is then defined as follows:
\begin{equation}
    attr_{j}(\boldsymbol{x}) = (\boldsymbol{x}_{j} - \boldsymbol{x}^{\prime}_{j}) \int_{\alpha=0}^{1} \frac{\partial f(\boldsymbol{x}^{\prime} + \alpha (\boldsymbol{x} - \boldsymbol{x}^{\prime}))}{\partial \boldsymbol{x}_{j}}d\alpha
\end{equation}
This integral is computed numerically by sampling the gradient evenly along the path. For our neural network ensemble, we average the integrated gradient attributions obtained from individual ensemble members to obtain a mean $attr_{j}$ for each feature.
\subsection{Transforming sensor signals into neural network inputs}
We compare NARX models with different combinations and transformations of the sensor data in their state variables $\mathbf{x}_{m-i}$ (Equation 2) and evaluate their accuracy.

Models that include the low frequency signals $p_{\text{cc}}, \dot m_{H2}, \dot m_{O2}, T_{H2}$ and $T_{O2}$ in their $\mathbf{x}_{m-i}$ vectors use the average value of the signal between $t_{m} - 50i$ and $t_{m} - 50(i+1)$. For the high frequency pressure und OH* signals, $p'$ and $I'$, we use transformations from the literature that are known to capture instability-relevant information as described next. The current $p'_{\text{rms}}$ is a natural candidate which is included in all models that incorporate pressure information. Sengupta \textit{et al} \cite{sengupta2020bayesian} and McCartney \textit{et al} \cite{mccartney2021reducing} have shown that before a combustor transitions to instability, the power spectral density shows measurable changes which can be used as a precursor. We estimate the spectral density of the $p'$ ($p'_{\text{FFT}}$) and $I'$ ($I'_{\text{FFT}}$) time series by applying Welch's method \cite{welch1967use} to a segment of the signals between $t_{m} - 50i$ ms and $t_{m} - 50(i+1)$ ms.  Welch's method divides the signal into overlapping segments and applies the Fast Fourier Transform (FFT) before averaging the result, leading to lower frequency resolution but a higher signal-to-noise ratio. In this study, a Hann window of length 30 with 50\% overlap is used for both $p'$ and $I'$. 

Kobayashi \cite{Kobayashi2019} applied machine learning to the the ordinal partition transition network of the $p'$ and $I'$ timeseries to detect instabilities early. The ordinal partition transition networks for an $m$-dimensional time series are expressed by a weighted adjacency matrix $W$ consisting of
$W_{ij} = P(\pi_i \longrightarrow \pi_j )$, $i, j \in [1, 2^m]$, $\sum W_{ij} = 1$, where $P(\pi_i \longrightarrow \pi_j )$ is
the observed probability from the $i$th- to $j$th-order transition patterns. In this study, the order patterns are $\pi_1$, $\pi_2$, $\pi_3$, and $\pi_4$ which account for all combinations of the signs of the increments for our two-dimensional time series $[p'(t), I'(t)]$. The probability distribution of the corresponding 16-dimensional transition patterns $W$ computed from the segment of the signals between $t_{m} - 50i$ and $t_{m} - 50(i+1)$ are used in the state variable vector. 

We also include a naive baseline model $p'_{\text{rms}}(t_m+\Delta t) = p'_{\text{rms}}(t_m)$ for comparison. This model, by design, is incapable of forecasting any future instabilities but, because large portions of the run consist of steady operation, its predictions have a reasonably good root mean squared error (RMSE). This model exists to provide a baseline RMSE and any useful forecast needs to be significantly more accurate than this.
\subsection{Train-test-validation split, hyperparameters, performance metrics}
We use a dataset of five experimental runs: three that are 50 seconds long and two that are 90 seconds long. Each run has two occurrences of the 1T instability, where an instability event is defined as being when the peak-to-peak amplitude of the pressure fluctuations exceeds 5\% of the chamber static pressure.

To avoid data leakage, one experimental run is exclusively reserved for hyperparameter tuning. The timeseries is split into contiguous 250 millisecond blocks and every fifth block is used to evaluate the negative log-likelihood, which is the metric we use for tuning the BNN hyperparameters. The hyperparameters are the network depth, layer width, noise $\sigma_{\epsilon}$, the learning rate for the ADAM optimizer, the number of epochs and the number of neural networks in the ensemble. For simplicity, and to facilitate comparison between models, the same hyperparameters are chosen for all models. This does not compromise the performance of any particular model, since increasing the key hyperparameters, width and depth, benefits all models. We choose a 3 hidden layer MLP with 100 units in each layer because larger networks show only minimal gains in performance. A learning rate of $2\times 10^{-4}$ is found to be optimal for convergence. The aleatoric noise parameter $\sigma_{\epsilon}$ is set to 0.07. In anchored ensembling, we train all members of the ensemble until convergence so the number of epochs was set to 512, which is found sufficient for all models. Converged estimates of the test data log-likelihood are obtained using 25 neural networks per ensemble.

To evaluate model performance, leave-one-out cross validation (LOOCV) is performed on the remaining four runs. Time series data involves strong temporal correlations, so assigning training and test data points randomly is dishonest because then the two sets would be highly similar. The test set must therefore be a completely independent experimental run. Additionally, LOOCV is necessary because with only four runs, firm conclusions cannot be drawn based on performance metrics for one particular test-train split, since any claimed efficacy or differences between models could be caused by chance. We report average performance metrics across the four possible splits.

To evaluate model accuracy, we compute the root mean squared error (RMSE) of our pressure amplitude forecast for the entire test run (full RMSE). We also calculate this on a subset of the test run comprising the two segments of the timeseries one second before and after a transition to instability event (transition RMSE). The RMSE is computed separately on this subset because our primary interest is in forecasting instabilities and the relative merits of different models become clearer on this subset where there is a dramatic change in the amplitude. As a measure of the quality of uncertainty, we report mean negative log-likelihood per test data point.

\section{Results}
The models were trained on a laptop with 16 GB RAM, an Intel i7-10870H CPU and an NVIDIA RTX 2070 GPU. Each NARX model took $\sim 2.5$ hours to train. The inference time of the ensemble averaged around $10$ milliseconds per sample, indicating that the algorithm is suitable as a real-time diagnostic tool. 

Table 1 shows the average prediction RMSEs for the NARX models computed during cross-validation across the four runs. We note, firstly, that the NARX models outperform the naive baseline, which means that the data contains a predictable signal that can be extracted. We also observe that the simple models, which use only the two operating point parameters $p_{\text{cc}}$ and ROF as inputs, perform well. This is unsurprising because we know that the 1T instabilities are consistently triggered when the operating point at $p_{\text{cc}} = 80$ bar and ROF $=6$ is approached. This means that knowing the current operating point and future flow rate control signals is informative. All transitions to instability in the test runs differ from each other, however, even though though they have identical control trajectories. The cross-validation shows unambiguously that the data from the additional instrumentation improves the prediction of future pressure fluctuations. Models that include the dynamic pressure spectrum $p'_{\text{FFT}}$, radiation intensity spectrum $I'_{\text{FFT}}$, current amplitude $p'_{\text{rms}}$, ordinal network transition probabilities, $W$, $T_{O2}$, $T_{H2}$, $p_{O2}$ or $p_{H2}$, have better prediction accuracies, especially on the transition subset. Including higher order dependencies on state variables also provides a slight boost in accuracy. Interestingly, the dynamic pressure spectrum appears more informative than the spectrum of the OH* radiation intensity. This could be due to the fact that a single optical probe observes only a small volume of the chamber, while the pressure field integrates information across the whole chamber.

\begin{table}[h!]
\centering
\begin{tabular}{||c c c c c||} 
 \hline
 State variable inputs $\mathbf{x}$ & Order & Full RMSE & Transition RMSE & NLL\\ [0.5ex] 
 \hline\hline
 Baseline & -- & 0.1552 & 0.2134 & --\\
 $p_{\text{cc}}$, ROF & 0 & 0.1289 & 0.1630 & -1.622\\ 
 $p_{\text{cc}}$, ROF & 2 & 0.1202 & 0.1576 & -1.631\\
 $p'_{\text{rms}}$, $p'_{\text{FFT}}$ & 0 & 0.1197 & 0.1546 & -1.5752\\ 
 $p'_{\text{rms}}$, $p'_{\text{FFT}}$ & 2 & 0.1169 & 0.1508 & -1.5467\\
 $p'_{\text{rms}}$, $I'_{\text{FFT}}$ & 0 & 0.1255 & 0.1612 & -1.5154\\ 
 $p'_{\text{rms}}$, $W$ & 0 & 0.1241 & 0.1566 & -1.5801\\
 $T_{O2}, T_{H2}, p_{O2}, p_{H2}$ & 0 & 0.1243 & 0.1578 & -1.5948\\
 $p_{\text{cc}}$, ROF, $p'_{\text{rms}}$, $p'_{\text{FFT}}$, $T_{O2}, T_{H2}, p_{O2}, p_{H2}$ & 0 & 0.1048 & 0.1407 & -1.5632\\
 $p_{\text{cc}}$, ROF, $p'_{\text{rms}}$, $p'_{\text{FFT}}$, $T_{O2}, T_{H2}, p_{O2}, p_{H2}$ & 2 & 0.1005 & 0.1311 & -1.5350\\[1ex] 
 \hline
\end{tabular}
\caption{Cross-validation RMSEs (in bar) for entire runs (Full RMSE) and the transition subset of the run (Transition RMSE) and mean negative log-likelihoods (NLL) for different NARX models. Lower RMSEs and NLLs are better.}
\label{table:1}
\end{table}

Table 1 also reports the mean negative log-likelihood (NLL) per datapoint, averaged across the four cross-validations. The mean negative log-likelihoods are fairly low for most of our models, implying that very few observations lie far outside the uncertainty bounds of the forecast (Figure 4). This is also confirmed by the percentage of data points in the test set within the $\pm 1$ s.d. and $\pm 3$ s.d. bounds, which are roughly in line with our Gaussian assumptions (68\% within 1 s.d., 99.7\% within 3 s.d.). Additionally, though models using more features have higher prediction accuracies, they have slightly worse NLLs in some cases because the presence of additional features results in higher epistemic uncertainties.

\begin{figure}
\centering
\includegraphics[scale = 0.155]{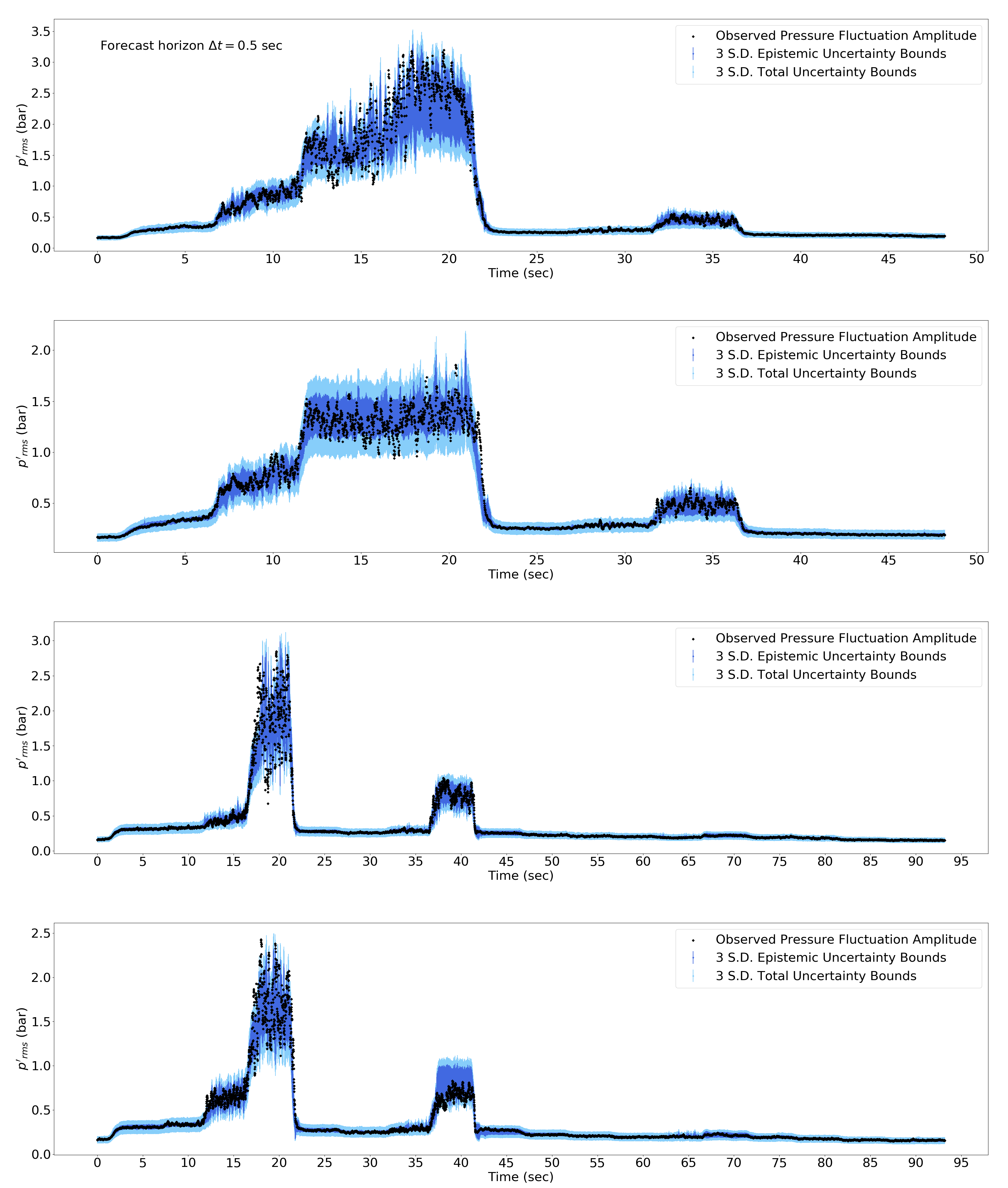}
\caption{Forecast $p'_{\text{rms}}$ values for the four experimental runs based on sensor data prior to $t - 0.5$ sec, superimposed on observed $p'_{\text{rms}}$ values at time $t$. The model was trained on the other three runs in each case. This model used $p'_{\text{FFT}}$ and $p'_{\text{rms}}$ as state variables and order $h=2$. Actual observations are shown as black dots, 3 S.D. total uncertainty bounds for the forecast are in light blue and 3 S.D. epistemic uncertainty bounds are in deep blue.}
\label{run4_good}
\end{figure}

\begin{figure}
\centering
\includegraphics[scale = 0.155]{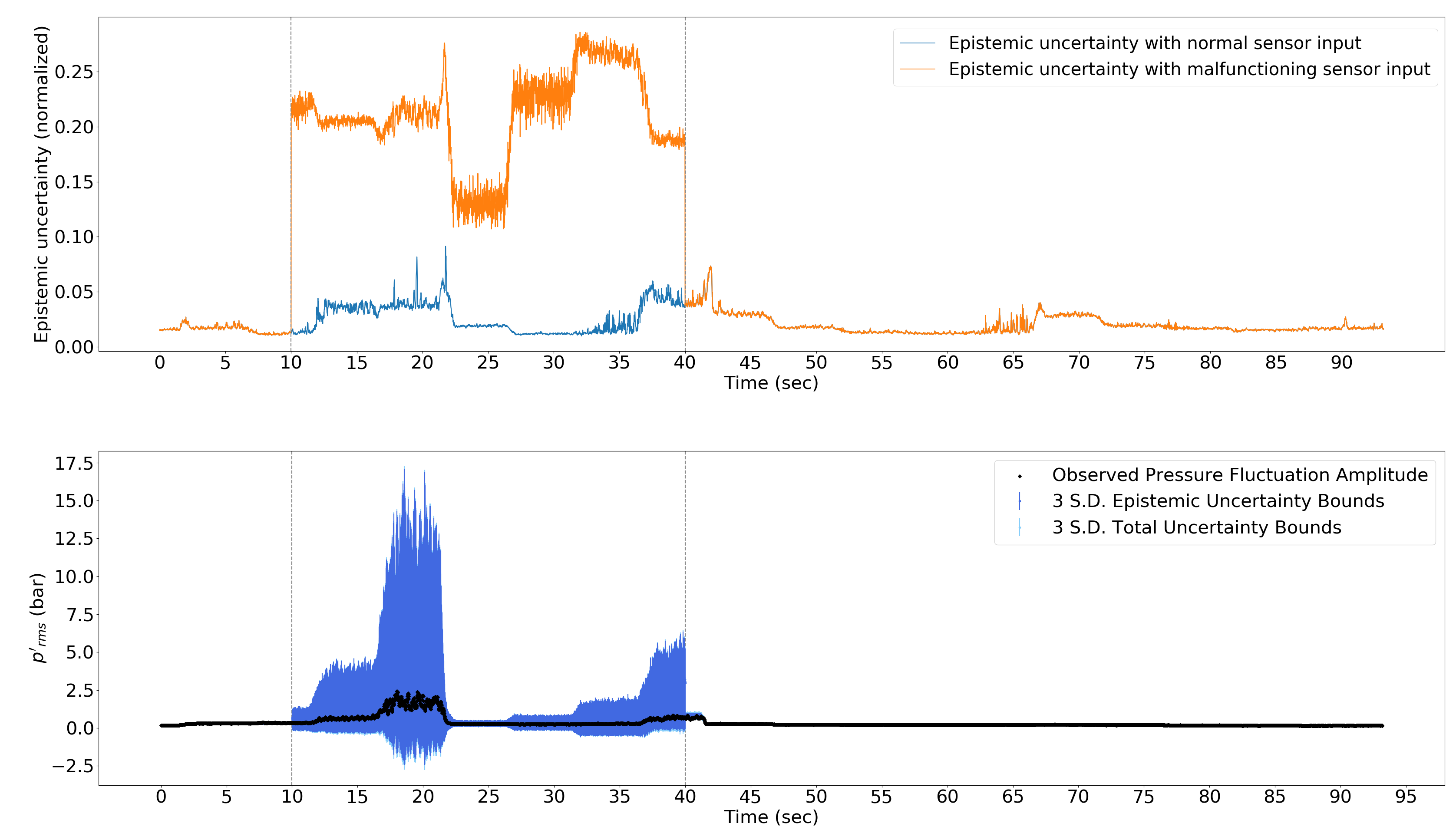}
\caption{Bayesian Neural Network epistemic uncertainty (top) and $p'_{\text{rms}}$ forecasts (bottom) for experimental run no. 4 where a fibre optical probe failure was simulated between $t = 10$ sec and $t = 40$ sec (marked by two grey dotted lines). The model used $I'_{\text{FFT}}, p'_{\text{rms}}$ as input state variables and order = 0. Actual observations are shown as black dots, 3 S.D. total uncertainty bounds for the forecast are represented by light blue bars and 3 S.D. epistemic uncertainty bounds are in deep blue.}
\label{run4_bad}
\end{figure}

To test the robustness of the Bayesian model against out-of-distribution inputs, we simulate a sensor failure event. This is shown in Figure \ref{run4_bad}. To simulate an optical probe icing event between 10 sec and 40 sec, the original photomultiplier signal is replaced by white noise with mean 0.05 and amplitude 0.015. During this simulated probe failure, the epistemic uncertainty of the Bayesian neural network greatly increases, indicating that the inputs are highly dissimilar to conditions experienced during training. Predictions remain reasonable despite the uninformative optical signal, although they become much less accurate.

\begin{figure}
\centering
\includegraphics[scale = 0.275]{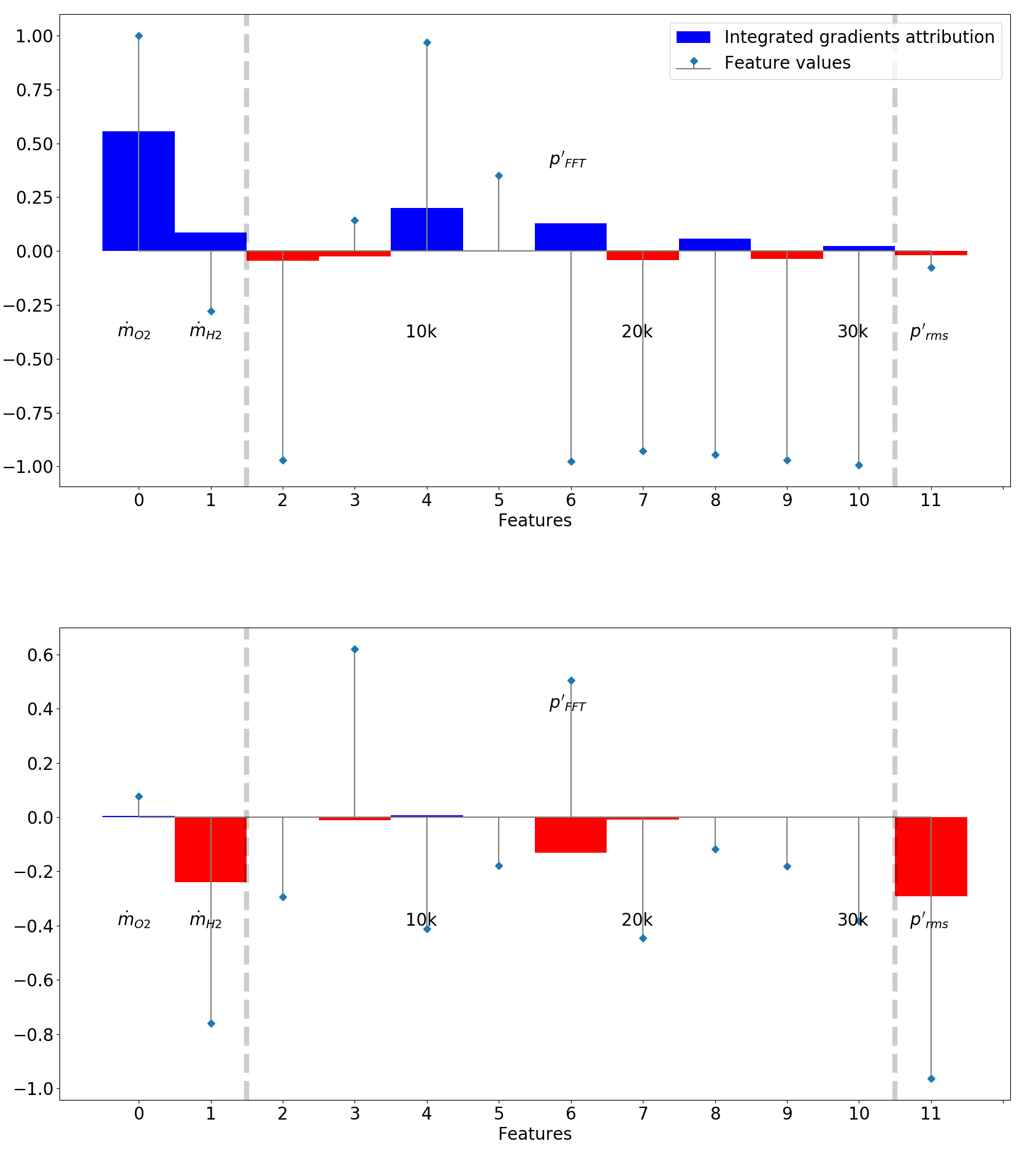}
\caption{Integrated gradient attributions and input feature values (normalized units) for two datapoints from experimental run no. 4 at t = 16.0 seconds (top, immediately preceding an instability) and t = 60.0 seconds (bottom, stable). The model used $p'_{\text{FFT}}$ and $p'_{\text{rms}}$ as state variables and order = 0. The blue dots show the normalized values of the features for this particular datapoint and the bars represent the integrated gradient attributions for each feature. For example, feature 6 (spectral power around the 16.66 kHz frequency) in the top plot has a large negative value and a positive integrated gradients attribution, meaning that the network has learned that low power in this frequency band is a potential indicator of instability.}
\label{integrad}
\end{figure}

We use the technique of integrated gradients to produce feature-level attribution plots, which can tell us how much a particular predictor influenced the prediction for a particular input. For example, the attribution plots in Figure \ref{integrad} shows this technique being applied to two data points. The first one is a data point just prior to the first instability event in a run, where a large jump in the dynamic pressure fluctuation amplitude is forecast by the model. This model uses $p'_{\text{FFT}}$, $p'_{\text{rms}}$ and the flowrate control signals as input. The most prominent feature is the large positive attribution for the oxygen flowrate control signal $\dot m_{O2}$, whose feature value is also large for this datapoint. This is expected, because a large $\dot m_{O2}$ would increase the $p_{\text{cc}}$ and ROF, bringing the combustor closer to instability. The prediction is also modulated by contributions from the $p'_{\text{FFT}}$ features. This datapoint has a concentration of power around the 10 kHz spectrum and low power in the mid to higher frequencies. There is a positive contribution from the 10 kHz frequency component of the spectrum, which is the frequency of the 1T mode, as well as from the 16.66k Hz frequency. This shows that the model has learned that an increased concentration of acoustic power around the frequency of the 1T mode and away from the mid-frequencies is an indication that the system is close to instability. The second data point is from stable operation. The attributions in this case have mostly negative values because the prediction of the model is strongly negative. The current pressure fluctuation amplitude $p'_{\text{rms}}$ is close to $-1.0$ in this case and naturally has a large negative attribution. The $\dot m_{H2}$ input (here $-0.78$) also contributes negatively because low ROFs are stable. Finally, the high concentration of power in the mid-frequencies also tells the network that this operating point will stay stable.


\section{Conclusions}

In this paper, we present the application of an uncertainty-aware and interpretable machine learning algorithm to forecast instabilities using multimodal sensor data in a realistic, thermoacoustically unstable rocket thrust chamber. We predict the amplitude of dynamic pressure fluctuations $500$ ms in advance using temperature, static pressure, high-frequency dynamic pressure and OH* radiation intensity data recorded during test runs performed with the cryogenic rocket thrust chamber BKD. We perform a rigorous cross-validation of the autoregressive Bayesian neural network and find that our models are able to anticipate all the instability events with varying accuracy ($25-40\%$ lower RMSEs on the transition subset compared to baseline), depending on the combination of inputs chosen. The inclusion of longer histories and additional sensor information, in general, boosts predictive accuracy. Models that use the dynamic pressure signal, in particular, show a measurable improvement in the forecast, confirming the findings of the literature on this topic. 

A key benefit of our Bayesian approach is the predictive uncertainty. In safety-critical machines such as rocket engines, a diagnostic tool must be robust to unexpected events such as sensor failures. We compute the log-likelihood of the test data given model predictions and find that uncertainties are well-characterized. We also simulate a sensor failure event in a model whose inputs included the spectrum of the fibre-optical probe signal and demonstrate the robustness of Bayesian algorithms. Integrated Gradients helps us identify important features and how they influence a particular prediction of the neural network.

Future work will focus on exploring the effectiveness of these tools on DLR's other experimental datasets, such as the LOX-methane tests, which also displayed injection-coupled acoustic instabilities \cite{klein2020injector}. We will also explore the generalizability of data-driven diagnostics across multiple datasets and combustors, using approaches that utilise transfer learning or domain invariance. Hardware upgrades are currently being undertaken at the test facility that will allow the implementation of these methods into the engine control algorithm \cite{waxenegger2020hardware}.

\section{Acknowledgements}
The authors would like to thank the crew of the P8 test bench and also Stefan Gröning for preparing and performing the test runs. Furthermore, we thank Wolfgang Armbruster for helping us process and understand the data. Ushnish Sengupta is an Early Stage Researcher within the MAGISTER consortium which receives funding from the European Union’s Horizon 2020 research and innovation programme under the Marie Skłodowska-Curie grant agreement No 766264. 

 \bibliographystyle{elsarticle-num} 
 \bibliography{cas-refs}





\end{document}